\documentclass{elsart}

\usepackage{epsf}

\def \lket {\left|}
\def \rket {\right\rangle}
\newcommand{\ket}[1]{\lket #1\rket}
\newcommand{\comment}[1]{}
\def \cin {\!\!\in\!\!}
\def \cnotin {\!\!\notin\!\!}
\def \Qspace {l_2(Q)}

\def\bbbz{{\mathchoice {\hbox{$\mathsf\textstyle Z\kern-0.4em Z$}}
{\hbox{$\mathsf\textstyle Z\kern-0.4em Z$}}
{\hbox{$\mathsf\scriptstyle Z\kern-0.3em Z$}}
{\hbox{$\mathsf\scriptscriptstyle Z\kern-0.2em Z$}}}}

\begin{document}
\begin{frontmatter}

\title{Exact results for accepting probabilities of quantum automata}

\author{Andris Ambainis\thanksref{tref1}}
\address{School of Mathematics,
Institute for Advanced Study, Princeton, NJ 08540}
\ead{ambainis@ias.edu}
\thanks[tref1]{Supported by NSF Grant CCR-9987845 and the State of
New Jersey. Part of this work done at University of California,
Berkeley, supported by Berkeley Fellowship for Graduate Studies, Microsoft 
Research Fellowship and and NSF Grant CCR-9800024.}

\author{Arnolds Kikusts\thanksref{tref2}}
\address{Institute of Mathematics and Computer Science,
 University of Latvia, Rai\c na bulv. 29, R\=\i ga,Latvia}
\ead{arnolds@usa.com}
\thanks[tref2]{Research supported by Grant No.01.0354 from the
Latvia Council of Science and European Commission,
contract IST-1999-11234.}

\maketitle

\begin{abstract}
One of the properties of the Kondacs-Watrous model of quantum finite automata
(QFA) is that the probability of the correct answer for a QFA cannot
be amplified arbitrarily. In this paper, we determine the
maximum probabilities achieved by QFAs for several languages.
In particular, we show that any language that is not recognized by an
RFA (reversible finite automaton) can be recognized by a QFA with
probability at most $0.7726...$.
\end{abstract}

\begin{keyword}
quantum computation, finite automata, quantum measurement.
\end{keyword}

\end{frontmatter}

\section{Introduction}

A quantum finite automaton (QFA) is a model for a quantum
computer with a finite memory.
QFAs can recognize the same languages as classical finite automata but
they can be exponentially more space efficient than 
their classical counterparts \cite{AF 98}.

To recognize an arbitrary regular language, QFAs need to be able
to perform general measurements after reading every input symbol,
as in \cite{AW 01,C 01,P 99}. 
If we restrict QFAs to unitary evolution and one measurement at
the end of computation (which might be easier to implement experimentally),
their power decreases considerably.
Namely \cite{CM 97,BP 99}, they can only recognize
the languages recognized by permutation automata,
a classical model in which the transitions between the states
have to be fully reversible.

Similar decreases of the computational power have been
observed in several other contexts. 
Quantum error correction is possible if we have a supply 
of quantum bits initialized to $\ket{0}$ at any moment
of computation (see chapter 10 of \cite{NC 00}).
Yet, if the number of quantum bits is fixed and it is not allowed 
to re-initialize them by measurements, error correction becomes difficult \cite{ABIN 96}.
Simulating a probabilistic Turing machine by a quantum Turing 
machine is trivial if we allow to measure and reinitialize qubits but
quite difficult if the number of qubits is fixed and they cannot be reinitialized
\cite{W 98}.

Thus, the availability of measurements is very important for
quantum automata.
What happens if the measurements are allowed but restricted?
How can we use the measurements of a restricted form to enhance
the abilities of quantum automata?
Can quantum effects be used to recognize languages
that are not recognizable by classical automata with the
same reversibility requirements?

In this paper, we look at those questions for 
``measure-many" QFA model by Kondacs and Watrous \cite{KW 97}.
This model allows intermediate measurements during the 
computation but these measurements
have to be of a restricted type. More specifically, they can have
3 outcomes: ``accept'', ``reject'', ``don't halt'' and if
one gets ``accept'' or ``reject'', the computation ends and this 
is the result of computation.
The reason for allowing measurements of this type was that
the states of a QFA then have a simple description
of the form $(\ket{\psi}, p_a, p_r)$ where $p_a$ is the probability
that the QFA has accepted, $p_r$ is the probability that the QFA
has rejected and $\ket{\psi}$ is the remaining state if the automaton
has not accepted or rejected.
Allowing more general measurements would make the remaining
state a mixed state $\rho$ instead of a pure state $\ket{\psi}$.
Having a mixed state as the current state of a QFA is 
very reasonable physically but the mathematical apparatus for
handling pure states is simpler than one for mixed states.


For this model, it is known that \cite{AF 98}
\begin{itemize}
\item
Any language recognizable by a 
QFA\footnote{For the rest of this paper, we will refer to ``measure-many''
QFAs as simply QFAs because this is the only model considered in
this paper.} 
with a probability $7/9+\epsilon$,
$\epsilon>0$ is recognizable by a reversible finite automaton 
(RFA).

\item
The language $a^{*} b^{*}$ can be recognized
with probability $0.6822..$ but cannot be recognized by an RFA.
\end{itemize}

\comment{Thus, the class of languages recognizable
with probability $0.6822...$ is not the same as the class of languages
recognizable with probability $7/9+\epsilon$.

In almost any other computational model, the accepting probability
can be increased by repeating the computation in parallel and,
usually, this property is considered completely obvious.
The above results showed
that this is not the case for QFAs. 
This is caused by the fact that the Kondacs-Watrous model  
of QFAs combines a reversible (unitary transformation) component
with a non-reversible component (measurements).}

Thus, the quantum automata in this model have an advantage over
their classical counterparts (RFAs) with the same reversibility requirements
but this advantage only allows to recognize languages with
probabilities at most 7/9, not $1-\epsilon$ with arbitrary $\epsilon>0$.
This is a quite unusual property because, 
in almost any other computational model, the accepting probability
can be increased by repeating the computation in parallel. 
As we see, this is not the case for QFAs. 


In this paper, we develop a method for determining
the maximum probability with which a QFA can recognize a 
given language. Our method is based on the quantum counterpart
of classification of states of a Markov chain into
ergodic and transient states \cite{KS 76}. 
We use this classification of states to transform the
problem of determining the maximum accepting probability
of a QFA into a quadratic optimization problem.
Then, we solve this problem (analytically in simpler cases, 
by computer in more difficult cases).

Compared to previous work, our new method has two advantages. 
First, it gives a systematic way of calculating the maximum accepting 
probabilities. Second, solving the optimization
problems usually gives the maximum probability exactly. 
Most of previous work \cite{AF 98,ABFK 99} used approaches
depending on the language and required two different methods:
one for bounding the probability from below, another for
bounding it from above. Often, using two different approaches
gave an upper and a lower bound with a gap between them
(like $0.6822...$ vs. $7/9+\epsilon$ mentioned above).
With the new approach, we are able to close those gaps.

We use our method to calculate the maximum accepting probabilities
for a variety of languages (and classes of languages).

First, we construct a quadratic optimization problem
for the maximum accepting probability by a QFA of
a language that is not recognizable by an RFA.
Solving the problem gives the probability $(52+4\sqrt{7})/81=0.7726...$.
This probability can be achieved for the language $a^{+}$
in the two-letter alphabet $\{a, b\}$
but no language that is no recognizable by a RFA can be recognized
with a higher probability.  
This improves the $7/9+\epsilon$
result of \cite{AF 98}.

This result can be phrased in a more general way.
Namely, we can find the property of a language which makes it impossible
to recognize the language by an RFA.
This property can be nicely stated in the form of
the minimal deterministic automaton containing a fragment 
of a certain form.

We call such a fragment a ``non-reversible construction".
It turns out that there are many different ``non-reversible
constructions'' and they have different influence on the accepting
probability. The one contained in the $a^{+}$ language
makes the language not recognizable by an RFA but the language
is still recognizable by a QFA with probability $0.7726...$.
In contrast, some constructions analyzed in \cite{BP 99,AKV 01}
make the language not recognizable with probability $1/2+\epsilon$
for any $\epsilon>0$.

In the rest of this paper, we look at different ``non-reversible constructions"
and their effects on the accepting probabilities of QFAs.
We consider three constructions: ``two cycles in a row'',
``$k$ cycles in parallel'' and a variant of the $a^+$ construction.
The best probabilities with which one can recognize languages
containing these constructions are $0.6894...$, $k/(2k-1)$ and $0.7324...$,
respectively. 

The solution of the optimization problem for ``two cycles in a row''
gives a new QFA for the language $a^* b^*$ that recognizes
it with probability $0.6894...$, improving the result of 
\cite{AF 98}. Again, using the solution of the optimization 
problem gives a better QFA that was previously missed because
of disregarding some parameters.

\section{Preliminaries}

\subsection{Quantum automata}


We define the Kondacs-Watrous (``measure-many'') model
of QFAs \cite{KW 97}.

\begin{def}
\label{def1}
A QFA is a tuple
$M=(Q;\Sigma ;V ;q_{0};Q_{acc};Q_{rej})$ where $Q$ is a finite set
of states, $\Sigma $ is an input alphabet, $V$ is a transition function
(explained below),
$q_{0}\cin Q$ is a starting state, and $Q_{acc}\subseteq Q$
and $Q_{rej}\subseteq Q$
are sets of accepting and rejecting states
($Q_{acc}\cap Q_{rej}=\emptyset$).
The states in $Q_{acc}$ and $Q_{rej}$,
are called {\em halting states} and
the states in $Q_{non}=Q-(Q_{acc}\cup Q_{rej})$ are called
{\em non halting states}.
\end{def}

{\bf States of $M$.} 
The state of $M$ can be any superposition of states in $Q$
(i. e., any linear combination of them with complex coefficients). 
We use $\ket{q}$ to denote the superposition consisting
of state $q$ only.
$l_2(Q)$ denotes the linear space consisting of all superpositions, with
$l_2$-distance on this linear space. 

{\bf Endmarkers.} 
Let $\kappa$ and $\$$ be symbols that do not belong to $\Sigma$.
We use $\kappa$ and $\$$ as the left and the right endmarker,
respectively. We call $\Gamma = \Sigma \cup \{\kappa ;\$\}$
the {\em working alphabet} of $M$.

{\bf Transition function.}
The transition function $V$ is a mapping from $\Gamma\times \Qspace$
to $\Qspace$ such that, for every $a\cin\Gamma$, the function
$V_a:\Qspace\rightarrow\Qspace$ defined by $V_a(x)=V(a, x)$ is a 
unitary transformation (a linear transformation on $l_2(Q)$ that
preserves $l_2$ norm).

{\bf Computation.}
The computation of a QFA starts in the superposition $\ket{q_{0}}$.
Then transformations corresponding to the left endmarker $\kappa$,
the letters of the input word $x$ and the right endmarker $\$$ are
applied. The transformation corresponding to $a\cin \Gamma$ consists
of two steps.

1. First, $V_{a}$ is applied. The new superposition $\psi^{\prime}$
is $V_{a}(\psi)$ where $\psi$ is the superposition before this step.

2. Then, $\psi^{\prime}$ is observed with respect to 
$E_{acc}, E_{rej}, E_{non}$ where
$E_{acc}=span\{|q\rangle:q\cin Q_{acc}\}$,
$E_{rej}=span\{|q\rangle :q\cin Q_{rej}\}$,
$E_{non}=span\{|q\rangle :q\cin Q_{non}\}$.
It means that if the system's state before the measurement was
$$\psi' = \sum_{q_i\in Q_{acc}} \alpha_i \ket{q_i} +
\sum_{q_j\in Q_{rej}} \beta_j \ket{q_j} +
\sum_{q_k\in Q_{non}} \gamma_k \ket{q_k}$$
then the measurement accepts $\psi'$ with probability $p_a=\Sigma\alpha_i^2$,
rejects with probability $p_r=\Sigma\beta_j^2$ and
continues the computation (applies transformations 
corresponding to next letters) with probability $p_c=\Sigma\gamma_k^2$
with the system having the (normalized) state $\frac{\psi}{\|\psi\|}$ 
where $\psi=\Sigma\gamma_k\ket{q_k}$.

We regard these two transformations as reading a letter $a$.

\comment{
{\bf Unnormalized states.}
Normalization (replacing $\psi$ by $\frac{\psi}{\|\psi\|}$)
is needed to make the probabilities of accepting, rejecting and
non-halting after the next letter sum up to 1.
However, normalizing the state after every letter can make 
the notation quite messy. (For the state after $k$ letters, there
would be $k$ normalization factors $\frac{1}{\|\psi_1\|}$,
$\ldots$, $\frac{1}{\|\psi_k\|}$ - one for each letter!)

For this reason, we do not normalize the states in our proofs.
That is, we apply the next transformations to the unnormalized state
$\psi$ instead of the normalized $\frac{\psi}{\|\psi\|}$. 

There is a simple correspondence between unnormalized and normalized
states. If, at some point, the unnormalized state is $\psi$,
then the normalized state is $\frac{\psi}{\|\psi\|}$
and the probability that the computation has not stopped 
is $\|\psi\|^2$.
$p a=\Sigma\alpha i^2$ and $p r=\Sigma\beta i^2$ become
the probabilities that the computation has not halted before this
moment but accepts (rejects) at this step.}

{\bf Notation.}
We use $V'_a$ to denote the transformation consisting of
$V_a$ followed by projection to $E_{non}$.
This is the transformation mapping $\psi$ to the 
non-halting part of $V_a(\psi)$. 
We use $V_w'$ to denote the product of transformations
$V_w'=V_{a_n}'V_{a_{n-1}}'\dots V_{a_2}'V_{a_1}'$,
where $a_i$ is the $i$-th letter of the word $w$.

We also use $\psi_w$ to denote the (unnormalized) non-halting part of 
QFA's state after reading the left endmarker $\kappa$ and the
word $w\cin\Sigma^*$.
From the notation it follows that $\psi_w=V_{\kappa w}'(\ket{q_0})$.

{\bf Recognition of languages.}
We will say that an automaton recognizes a language $L$ with probability $p$
$(p>\frac{1}{2})$ if it accepts any word $x\cin L$ with
probability $\geq p$ and
rejects any word $x\cnotin L$ with probability $\geq p$.

\subsection{Useful lemmas}

For classical Markov chains, one can classify the states of a Markov
chain into {\em ergodic} sets and {\em transient} sets \cite{KS 76}.
If the Markov chain is in an ergodic set, it never leaves it.
If it is in a transient set, it leaves it with probability
$1-\epsilon$ for an arbitrary $\epsilon>0$ after
sufficiently many steps.

A quantum counterpart of a Markov chain is a quantum system
to which we repeatedly apply a transformation that depends 
on the current state of the system but does not depend on
previous states. In particular, it can be a QFA that repeatedly
reads the same word $x$. Then, the state after reading $x$ $k+1$ 
times depends on the state after reading $x$ $k$ times but
not on any of the states before that.
The next lemma gives the classification of states
for such QFAs.

\begin{lem}
\label{LemmaAF}
\cite{AF 98}
Let $x\in \Sigma^{+}$.
There are subspaces $E_1$, $E_2$ such that $E_{non}=E_1\oplus E_2$ and
\begin{enumerate}
\item[(i)]
If $\psi\in E_1$, then $V'_x(\psi)\in E_1$ and $\| V'_x(\psi)\|=\|\psi\|$,
\item[(ii)]
If $\psi\in E_2$, then $\| V'_{x^k}(\psi)\|\rightarrow 0$ when
$k\rightarrow\infty$.
\end{enumerate}
\end{lem}

Instead of ergodic and transient sets, we have subspaces $E_1$ and $E_2$.
The subspace $E_1$ is a counterpart of an ergodic set:
if the quantum 
process defined by repeated reading of $x$
is in a state $\psi\in E_1$, it stays in $E_1$.
$E_2$ is a counterpart of a transient set:
if the state is $\psi\in E_2$, $E_2$ is left (for an accepting
or rejecting state) with probability arbitrarily close to 1
after sufficiently many $x$'s.

In some of proofs 
we also use a generalization of Lemma \ref{LemmaAF}
to the case of two (or more) words $x$ and $y$:

\begin{lem}
\label{AKVLemma}
\cite{AKV 01}
Let $x,y\in \Sigma^{+}$.
There are subspaces $E_1$, $E_2$ such that $E_{non}=E_1\oplus E_2$ and
\begin{enumerate}
\item[(i)]
If $\psi\in E_1$, then $V'_x(\psi)\in E_1$ and $V'_y(\psi)\in E_1$
and $\|V'_x(\psi)\|=\|\psi\|$ and $\|V'_y(\psi)\|=\|\psi\|$,
\item[(ii)]
If $\psi\in E_2$, then for any $\epsilon >0$,
there exists $t\in (x|y)^*$ such that
$\| V'_t(\psi)\|<\epsilon$.
\end{enumerate}
\end{lem}

We also use a lemma from \cite{BV 97}.

\begin{lem}
\label{BVLemma}
\cite{BV 97}
If $\psi$ and $\phi$ are two quantum states and
$\|\psi-\phi\|<\varepsilon$ then the total variational distance
between probability distributions generated by the same
measurement on $\psi$ and $\phi$ is at most\footnote{The 
lemma in \cite{BV 97} has $4\varepsilon$ but 
it can be improved to $2\varepsilon$.} $2\varepsilon$.
\end{lem}

\section{QFAs vs. RFAs}

Ambainis and Freivalds \cite{AF 98} 
characterized the languages recognized
by RFAs as follows.

\begin{thm}
\cite{AF 98}
\label{AFTheorem}
Let $L$ be a language and $M$ be its minimal automaton.
$L$ is recognizable by a RFA if and only if
there is no $q_1, q_2, x$ such that
\begin{enumerate}
\item
$q_1\neq q_2$,
\item
If $M$ starts in the state $q_1$ and reads $x$,
it passes to $q_2$, 
\item
If $M$ starts in the state $q_2$ and reads $x$,
it passes to $q_2$, and
\item
$q_2$ is neither "all-accepting" state, nor "all-rejecting" state,
\end{enumerate}
\end{thm}

An RFA is a special case of a QFA that outputs the correct answer 
with probability 1. Thus, any language that does not contain
the construction of Theorem \ref{AFTheorem} can be recognized by
a QFA that always outputs the correct answer.
Ambainis and Freivalds \cite{AF 98} also showed the reverse of this:
any language $L$ with the minimal automaton
containing the construction of Theorem \ref{AFTheorem}
cannot be recognized by a QFA with probability $7/9+\epsilon$.

\begin{figure}
\begin{center}
\epsfxsize=3in
\hspace{0in}
\epsfbox{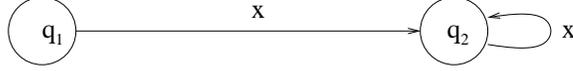}
\caption{\it ``The forbidden construction'' of Theorem \ref{AFTheorem}.}
\label{F1}
\end{center}
\end{figure}

We consider the question: what is the maximum probability of
correct answer than can be achieved by a QFA for a language
that cannot be recognized by an RFA?
The answer is:

\begin{thm}
\label{T1}
Let $L$ be a language and $M$ be its minimal automaton.
\begin{enumerate}
\item
If $M$ contains the construction of Theorem \ref{AFTheorem},
$L$ cannot be recognized by a 1-way QFA with 
probability more than $p=(52+4\sqrt{7})/81=0.7726...$.
\item
There is a language $L$ with the minimal automaton $M$ containing 
the construction of Theorem \ref{AFTheorem} that can be recognized
by a QFA with probability $p=(52+4\sqrt{7})/81=0.7726...$.
\end{enumerate}
\end{thm}

{\em Proof.}
We consider the following optimization problem.

{\bf Optimization problem 1.}
Find the maximum $p$ such that there is a finite dimensional
vector space $E_{opt}$, subspaces $E_a$, $E_r$ such that $E_a\perp E_r$,
vectors $v_1$, $v_2$ such that $v_1\perp v_2$ and $\|v_1+v_2\|= 1$
and probabilities $p_1$, $p_2$ such that $p_1+p_2=\|v_2\|^2$ and
\begin{enumerate}
\item
$\|P_a(v_1+v_2)\|^2\geq p$,
\item
$\|P_r(v_1)\|^2+p_2\geq p$,
\item
$p_2\leq 1-p$.
\end{enumerate}

We sketch the relation between a QFA recognizing $L$
and this optimization problem. 
Let $Q$ be a QFA recognizing $L$.
Let $p_{min}$ be the minimum probability of the correct
answer for $Q$, over all words.
We use $Q$ to construct an instance of the optimization problem
above with $p\geq p_{min}$.

Namely, we look at $Q$ reading an infinite (or very long finite)
sequence of letters $x$.
By Lemma \ref{LemmaAF}, we can decompose the starting state $\psi$
into 2 parts $\psi_1\in E_1$ and $\psi_2\in E_2$.
Define $v_1=\psi_1$ and $v_2=\psi_2$.
Let $p_1$ and $p_2$ be the probabilities of getting into
an accepting (for $p_1$) or rejecting (for $p_2$)
state while reading an infinite sequence of $x$'s 
starting from the state $v_2$.
The second part of Lemma \ref{LemmaAF} implies
that $p_1+p_2=\|v_2\|^2$.

Since $q_1$ and $q_2$ are different states of the minimal automaton $M$,
there is a word $y$ that is accepted in one of them 
but not in the other. Without loss of generality,
we assume that $y$ is accepted if $M$ is started in $q_1$
but not if $M$ is started in $q_2$.
Also, since $q_2$ is not an ``all-accepting'' state,
there must be a word $z$ that is rejected if
$M$ is started in the state $q_2$.

We choose $E_a$ and $E_r$ so that the square of the projection $P_a$ 
($P_r$) of a vector $v$ on $E_a$ ($E_r$) is equal to the accepting
(rejecting) probability of $Q$ if we run $Q$ on the starting state $v$
and input $y$ and the right endmarker $\$$.

Finally, we set $p$ equal to the $\inf$ of the set consisting
of the probabilities of correct answer of $Q$ on the words
$y$ and $x^i y$, $x^i z$ for all $i\in\bbbz$.

Then, Condition 1 of the optimization problem,
$\|P_a(v_1+v_2)\|^2\geq p$ is true because the word
$y$ must be accepted and the accepting probability for it is
exactly the square of the projection of the starting
state ($v_1+v_2$) to $P_a$.

Condition 2 follows from running $Q$ on a word
$x^i y$ for some large $i$. 
By Lemma \ref{LemmaAF}, if $i>k$ for some $k$,
$\|V'_{x^i}(v_2)\|\leq \epsilon$.
Also, $v_1$, $V'_{x}(v_1)$, $V'_{x^2}(v_1)$, $\ldots$
is an infinite sequence in a finite-dimensional space.
Therefore, it has a limit point and there are $i, j$, $i\geq k$
such that 
\[ \| V'_{x^j}(v_1)-V'_{x^{i+j}}(v_1)\|\leq\epsilon .\]
We have
\[ V'_{x^j}(v_1) - V'_{x^{i+j}}(v_1) = V'_{x^j} (v_1-V'_{x^i}(v_1)) .\]
Since $\|V'_x(\psi)\|=\|\psi\|$ for $\psi\in E_1$,
$\|V'_{x^j} (v_1-V'_{x^i}(v_1)) \| = \|v_1-V'_{x^i}(v_1) \|$ and we have 
\[ \| v_1-V'_{x^i}(v_1)\|\leq\epsilon .\]
Thus, reading $x^i$ has the following effect:
\begin{enumerate}
\item
$v_1$ gets mapped to a state 
that is at most $\epsilon$-away (in $l_2$ norm) from $v_1$,
\item
$v_2$ gets mapped to an accepting/rejecting state
and most $\epsilon$ fraction of it stays on the non-halting states. 
\end{enumerate}
Together, these two requirements mean that the state of $Q$ 
after reading $x^i$ is at most $2\epsilon$-away from $v_1$.
Also, the probabilities of $Q$ accepting and rejecting while
reading $x^i$ differ from $p_1$ and $p_2$
by at most $\epsilon$.

Let $p_{x^iy}$ be the probability of $Q$ rejecting $x^i y$.
Since reading $y$ in $q_2$ leads to a rejection, $x^i y$ must
be rejected and $p_{x^iy}\geq p$. 
The probability $p_{x^iy}$ consists of
two parts: the probability of rejection during $x^i$ and
the probability of rejection during $y$.
The first part differs from $p_2$ by at most $\epsilon$,
the second part differs from $\|P_r(v_1)\|^2$ by at most $4\epsilon$
(because the state of $Q$ when starting to read $y$ differs
from $v_1$ by at most $2\epsilon$ and, by Lemma \ref{BVLemma},
the accepting probabilities differ by at most twice that).
Therefore, 
\[  p_{x^iy}-5\epsilon \leq p_2+\|P_r(v_1)\|^2 \leq p_{x^i y}+5\epsilon
.\] 
Since $p_{x^iy}\geq p$, this implies 
$p-5\epsilon\leq p_2+\|P_r(v_1)\|^2$.
By appropriately choosing $i$, 
we can make this true for any $\epsilon>0$.
Therefore, we have $p\leq p_2+\|P_r(v_1)\|^2$
which is Condition 2.

Condition 3 is true by considering $x^i z$.
This word must be accepted with probability $p$.
Therefore, for any $i$, $Q$ can only reject during
$x^i$ with probability $1-p$ and $p_2\leq 1-p$.

This shows that no QFA can achieve a probability
of correct answer more than the solution of optimization
problem 1. It remains to solve this problem.

{\bf Solving Optimization problem 1.}

The key idea is to show that it is enough to consider 2-dimensional
instances of the problem.


Since $v_1\perp v_2$, the vectors $v_1, v_2, v_1+v_2$
form a right-angled triangle.
This means that $\|v_1\|=\cos\beta \|v_1+v_2\|=\cos\beta$,
$\|v_2\|=\sin\beta \|v_1+v_2\|=\sin\beta$ where
$\beta$ is the angle between $v_1$ and $v_1+v_2$.
Let $w_1$ and $w_2$ be the normalized versions
of $v_1$ and $v_2$: $w_1=\frac{v_1}{\|v_1\|}$,
$w_2=\frac{v_2}{\|v_2\|}$. 
Then, $v_1=\cos\beta w_1$ and $v_2=\sin\beta w_2$.

Consider the two-dimensional subspace spanned
by $P_a(w_1)$ and $P_r(w_1)$.
Since the accepting and the rejecting subspaces
$E_a$ and $E_r$ are orthogonal,
$P_a(w_1)$ and $P_r(w_1)$ are orthogonal.
Therefore, the vectors $w_a=\frac{P_a(w_1)}{\|P_a(w_1)\|}$ and
$w_r=\frac{P_r(w_1)}{\|P_r(w_1)\|}$ form an orthonormal basis.
We write the vectors $w_1$, $v_1$ and $v_1+v_2$ in this basis.
The vector $w_1$ is $(\cos \alpha, \sin\alpha)$
where $\alpha$ is the angle between $w_1$ and $w_a$.
The vector $v_1=\cos\beta w_1$ is equal to
$(\cos\beta \cos\alpha, \cos\beta \sin\alpha)$.

Next, we look at the vector $v_1+v_2$.
We fix $\alpha$, $\beta$ and $v_1$ and try to find the
$v_2$ which maximizes $p$ for the fixed $\alpha$, $\beta$ and $v_1$.
The only place where $v_2$ appears in the optimization problem 1
is $\|P_a(v_1+v_2)\|^2$ on the left hand side of Condition 1.
Therefore, we should find $v_2$ that maximizes $\|P_a(v_1+v_2)\|^2$.
We have two cases:
\begin{enumerate}
\item
$\alpha\geq \beta$.

The angle between $v_1+v_2$ and $w_a$ is at least
$\alpha-\beta$
(because the angle between $v_1$ and
$w_a$ is $\alpha$ and the angle between $v_1+v_2$
and $v_1$ is $\beta$).
Therefore, the projection of $v_1+v_2$ to $w_a$ is at
most $\cos(\alpha-\beta)$.
Since $w_r$ is a part of the rejecting subspace $E_r$,
this means that $\|P_a(v_1+v_2)\|^2\leq \cos^2(\alpha-\beta)$.
The maximum $\|P_a(v_1+v_2)\|=\cos^2(\alpha-\beta)$ 
is achieved if we put $v_1+v_2$ in the plane spanned by 
$w_a$ and $w_r$: $v_1+v_2= (\cos(\alpha-\beta), \sin(\alpha-\beta))$.

Next, we can rewrite Condition 3 of the optimization problem
as $1-p_2\geq p$.
Then, Conditions 1-3 together mean that 
\begin{equation}
\label{eq1}
p=\min(\|P_a(v_1+v_2)\|^2, \|P_r(v_1)\|^2+p_2, 1-p_2).
\end{equation}
To solve the optimization problem, we have to 
maximize (\ref{eq1}) subject to the conditions of the problem.
From the expressions for $v_1$ and $v_1+v_2$ above,
it follows that (\ref{eq1}) is equal to 
\begin{equation}
\label{eq2} 
p=\min(\cos^2(\alpha-\beta), \sin^2\alpha\cos^2\beta+p_2, 1-p_2) 
\end{equation}
First, we maximize $\min(\sin^2\alpha\cos^2\beta+p_2, 1-p_2)$.
The first term is increasing in $p_2$, the second is decreasing.
Therefore, the maximum is achieved when both become equal which
happens when $p_2=\frac{1-\sin^2\alpha\cos^2\beta}{2}$.
Then, both $\sin^2\alpha\cos^2\beta+p_2$ and $1-p_2$ are
$\frac{1+\sin^2\alpha\cos^2\beta}{2}$.
Now, we have to maximize
\begin{equation}
\label{eq3} 
p=\min\left(\cos^2(\alpha-\beta), 
\frac{1+\sin^2\alpha\cos^2\beta}{2}\right) .
\end{equation}
We first fix $\alpha-\beta$ and try to optimize the second term.
Since $\sin\alpha\cos\beta=\frac{\sin(\alpha+\beta)+\sin(\alpha-\beta)}{2}$
(a standard trigonometric identity),
it is maximized when $\alpha+\beta=\frac{\pi}{2}$ and
$\sin(\alpha+\beta)=1$.
Then, $\beta=\frac{\pi}{2}-\alpha$ and (\ref{eq3}) becomes
\begin{equation} 
\label{eq4}
p=\min\left(\sin^2 2\alpha, \frac{1+\sin^4\alpha}{2} \right) .
\end{equation}
The first term is increasing in $\alpha$, the second is decreasing.
The maximum is achieved when
\begin{equation}
\label{eq5}
 \sin^2 2\alpha=\frac{1+\sin^4 \alpha}{2} .
\end{equation}
The left hand side of (\ref{eq5}) is equal to $4\sin^2\alpha\cos^2\alpha=
4\sin^2\alpha(1-\sin^2\alpha)$. Therefore, if we denote $\sin^2\alpha$ by $y$,
(\ref{eq5}) becomes a quadratic equation in $y$:
\[ 4 y(1-y)=\frac{1+y^2}{2} .\]
Solving this equation gives $y=\frac{4+\sqrt{7}}{9}$
and $4y(1-y)=\frac{52+4\sqrt{7}}{81}=0.7726...$.
\item
$\alpha<\beta$.

We consider $\min(\|P_r(v_1)\|^2+p_2, 1-p_2)=
\min(\sin^2\alpha\cos^2\beta+p_2, 1-p_2)$.
Since the minimum of two quantities is at most 
their average, this is at most
\begin{equation}
\label{eq6} 
\frac{1+\sin^2\alpha \cos^2\beta}{2} .
\end{equation}
Since $\alpha<\beta$, we have $\sin\alpha<\sin\beta$ and 
(\ref{eq6}) is at most $\frac{1+\sin^2\beta\cos^2\beta}{2}$.
This is maximized by $\sin^2\beta=1/2$.
Then, we get $\frac{1+1/4}{2}=\frac{5}{8}$
which is less than $p=0.7726...$ which we got in the first case.
\end{enumerate}

This proves the first part of the theorem. \qed


{\bf Construction of a QFA.}

This part is proven by taking the solution 
of optimization problem 1 and using it to construct a QFA for
the language $a^+$ in a two-letter alphabet
$\{a, b\}$. The state $q_1$ is just the starting state of
the minimal automaton, $q_2$ is the state to which it gets after reading $a$,
$x=a$, $y$ is the empty word and $z=b$.

Let $\alpha$ be the solution of (\ref{eq5}).
Then, $\sin^2\alpha=(4+\sqrt{7})/9$,
$\cos^2\alpha = 1-\sin^2\alpha = (5-\sqrt{7})/9$,
$\cos 2\alpha=\cos^2\alpha-\sin^2\alpha=(1-2\sqrt{7})/9$,
$\cos^2 2\alpha=(1-2\sqrt{7})^2/81=(29-4\sqrt{7})/81$
and $\sin^2 2\alpha=1-\cos^2 2\alpha=(52+4\sqrt{7})/81$.
$\sin^2 2\alpha$ is the probability of correct answer for our QFA described below.

The QFA $M$ has 5 states: $ q_0,  q_1,  q_{acc}$, $q_{rej}$ and 
$q_{rej1}$.
$Q_{acc}=\{ q_{acc}\}$, $Q_{rej}=\{ q_{rej}, q_{rej1}\}$.
The initial state is $\sin\alpha |q_0\rangle+\cos\alpha |q_1\rangle$.
The transition function is 
\[ V_a(|q_0\rangle)=|q_0\rangle, V_a(|q_1\rangle)= 
\sqrt{\frac{1+\sin^2\alpha}{2}}|q_{acc}\rangle+
\frac{\cos\alpha}{\sqrt{2}}|q_{rej}\rangle,\]
\[ V_b(|q_0\rangle)=|q_{rej}\rangle, V_b(|q_1\rangle)=|q_{rej1}\rangle,\]
\[ V_{\$}(|q_0\rangle)=\sin\alpha |q_{acc}\rangle+\cos\alpha |q_{rej}\rangle,
V_{\$}(|q_1\rangle)=-\cos\alpha |q_{acc}\rangle+\sin\alpha |q_{rej}\rangle\]
To recognize $L$, $M$ must accept all words of the form $a^i$ for $i>0$
and reject the empty word and any word that contains the letter $b$.
\begin{enumerate}
\item
The empty word.

The only tranformation applied to the starting state is $V_{\$}$.
Therefore, the final superposition is
\[ V_{\$}(\sin\alpha |q_0\rangle+\cos\alpha |q_1\rangle)=
(\sin^2\alpha-\cos^2\alpha)|q_{acc}\rangle
+2\sin\alpha\cos\alpha|q_{rej}\rangle.\]
The amplitude of $|q_{rej}\rangle$ in the final superposition is 
$2\sin\alpha\cos\alpha=\sin 2\alpha$ and the word is rejected with a
probability $\sin^2 2\alpha=0.772...$.
\item
$a^i$ for $i>0$.

First, $V_a$ maps the $\cos|q_1\rangle$ component to 
\[ \cos\alpha \sqrt{\frac{1+\sin^2\alpha}{2}}|q_{acc}\rangle+
\frac{\cos^2\alpha}{\sqrt{2}}|q_{rej}\rangle.\] 
The probability of accepting at this point is 
$\cos^2 \alpha\frac{1+\sin^2\alpha}{2}$. 
The other component of the superposition,
$\sin\alpha|q_0\rangle$ stays unchanged until $V_{\$}$ maps it to 
\[ \sin^2\alpha|q_{acc}\rangle+\sin\alpha\cos\alpha|q_{rej}\rangle.\]
The probability of accepting at this point is $\sin^4\alpha$.
The total probability of accepting is
\[ 
\cos^2 \alpha\frac{1+\sin^2\alpha}{2} + \sin^4\alpha = 
(1-\sin^2 \alpha)\frac{1+\sin^2\alpha}{2} + \sin^4\alpha = 
\frac{1+\sin^4\alpha}{2} .
\] \
By equation (\ref{eq6}), this is equal to $\sin^2 2\alpha$.
\item
A word containing at least one $b$.

If $b$ is the first letter of the word, the entire superposition
is mapped to rejecting states and the word is rejected with probability 1.
Otherwise, the first letter is $a$, it maps $\cos\alpha|q_1\rangle$ to 
$\cos\alpha \sqrt{\frac{1+\sin^2\alpha}{2}}|q_{acc}\rangle+
\frac{\cos^2\alpha}{\sqrt{2}}|q_{rej}\rangle$.
The probability of accepting at this point is 
$\cos^2\alpha(1+\sin^2\alpha)/2= (1-\sin^2\alpha)(1+\sin^2\alpha)/2=
(1-\sin^4\alpha)/2$.
By equation (\ref{eq6}), this is the same as $1-\sin^2 2\alpha$.
After that, the remaining component ($\sin\alpha|q_0\rangle$) is not 
changed by next $a$s and mapped to a rejecting state by the first $b$.
Therefore, the total probability of accepting is also $1-\sin^2 2\alpha$
and the correct answer (rejection) is given with a probability 
$\sin^2 2\alpha$.
\end{enumerate}
\qed

\section{Non-reversible constructions}

We now look at fragments of the minimal automaton that imply
that a language cannot be recognized with probability more than $p$,
for some $p$. We call such fragments ``non-reversible constructions".
The simplest such construction is the one of Theorem \ref{AFTheorem}.
In this section, we present 3 other ``non-reversible constructions" 
that imply that a language can be recognized with probability at most
$0.7324...$, $0.6894...$ and $k/(2k-1)$.
This shows that different constructions are ``non-reversible"
to different extent. Comparing these 4 ``non-reversible'' constructions
helps to understand what makes one of them harder for
QFA (i.e., recognizable with worse probability of correct answer)

\subsection{``Two cycles in a row''}

The first construction comes from the language $a^* b^*$ considered
in Ambainis and Freivalds \cite{AF 98}. This language was the first 
example of a language that can be recognized by a QFA with some 
probability (0.6822...) but not with another ($7/9+\epsilon$).
We find the ``non-reversible'' construction for this language
and construct the QFA with the best possible accepting probability.

\begin{thm}
\label{TF4}
Let $L$ be a language and $M$ its minimal automaton.
\begin{enumerate}
\item
If $M$ contains states $q_1$, $q_2$ and $q_3$ such that,
for some words $x$ and $y$,
\begin{enumerate}
\item
if $M$ reads $x$ in the state $q_1$, it passes to $q_1$, 
\item
if $M$ reads $y$ in the state $q_1$, it passes to $q_2$,
\item
if $M$ reads $y$ in the state $q_2$, it passes to $q_2$, 
\item
if $M$ reads $x$ in the state $q_2$, it passes to $q_3$,
\item
if $M$ reads $x$ in the state $q_3$, it passes to $q_3$
\end{enumerate}
then $L$ cannot be recognized by a QFA
with probability more than $0.6894...$.
\item
The language $a^*b^*$ (the minimal automaton of which
contains the construction above) can be recognized by a QFA with 
probability $0.6894...$.
\end{enumerate}
\end{thm}

\noindent {\em Proof.} 
By a reduction 
to the following optimization problem.

\begin{figure}
\begin{center}
\epsfxsize=3in
\hspace{0in}
\epsfbox{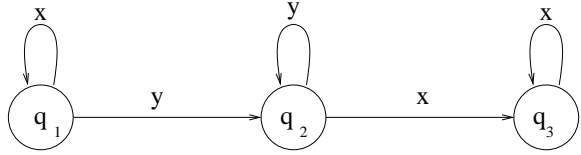}
\caption{\it ``The forbidden construction'' of Theorem \ref{TF4}.}
\label{F4}
\end{center}
\end{figure}

{\bf Optimization problem 2.}
Find the maximum $p$ such that there is 
a finite-dimensional space $E$,
subspaces $E_a$, $E_r$ such that $E=E_a\oplus E_r$,
vectors $v_1$, $v_2$ and $v_3$ and 
probabilities $p_{a_1}$, $p_{r_1}$, $p_{a_2}$, $p_{r_2}$ such that
\begin{enumerate}
\item
$\| v_1+v_2+v_3\|=1$,
\item
$v_1 \perp v_2$,
\item
$v_1+v_2+v_3 \perp v_2$,
\item
$v_1+v_2 \perp v_3$.
\item
$\|v_3\|^2=p_{a_1}+p_{r_1}$;
\item
$\|v_2\|^2=p_{a_2}+p_{r_2}$;
\item 
$\|P_a(v_1+v_2+v_3)\|^2 \geq p$;
\item
$\|P_a(v_1+v_2)\|^2+p_{a_1} \geq p$;
\item
$\|P_a(v_1)\|^2+p_{a_1}+p_{a_2} \leq 1-p$.
\end{enumerate}

%
%
%

We use a theorem from \cite{BP 99}.

\begin{thm}
\label{T13}
Let $L$ be a language and $M$ be its minimal automaton.
Assume that there is a word $x$ such that $M$  
contains states $q_1$, $q_2$ satisfying:
\begin{enumerate}
\item
$q_1\neq q_2$,
\item
If $M$ starts in the state $q_1$ and reads $x$,
it passes to $q_2$,
\item
If $M$ starts in the state $q_2$ and reads $x$,
it passes to $q_2$, and
\item
There is a word $y$ such that if M starts in $q_2$ and reads y, it passes to $q_1$,
\end{enumerate}
then $L$ cannot be recognized by any 1-way quantum finite automaton.
\end{thm}

Let $Q$ be a QFA recognizing $L$. 
Let $q_4$ be state where the minimal automaton $M$ goes if it reads $y$
in the state $q_3$.
In case when $q_2=q_4$ we get the forbidden construction of Theorem 
\ref{T13}.
In case when $q_2\neq q_4$ states $q_2$ and $q_4$ are different
states of the minimal automaton $M$. Therefore, 
there is a word $z$ that is accepted in one of them 
but not in the other. Without loss of generality,
we assume that $y$ is accepted if $M$ is started in $q_2$
but not if $M$ is started in $q_4$.

We choose $E_a$ so that the square of the projection $P_a$ 
of a vector $v$ on $E_a$ is equal to the accepting
probability of $Q$ if we run $Q$ on the starting state $v$
and input $yz$ and the right endmarker $\$$.

We use Lemma \ref{LemmaAF}.
Let $E_1^x$ be $E_1$ and $E_2^x$ be $E_2$ for word $x$  
and let $E_1^y$ be $E_y$ and $E_2^y$ be $E_y$ for word $y$.

Without loss of generality we can assume that $q_1$ is a 
starting state of $M$. 
Let $\psi_\kappa$ be the starting superposition for $Q$.
We can also assume that reading $x$ in this state
does not decrease the norm of this superposition.
We divide $\psi_\kappa$ into three parts: $v_1$, $v_2$ and $v_3$ so that
$v_1+v_2\in E_1^y$ and $v_3\in E_2^y$, 
$V_1\in E_1^x$ and $v_2\in E_2^x$. 
Due to $v_1+v_2+v_3$ is the starting superposition we have 
$||v_1+v_2+v_3||=1$(Condition 1).

Since $v_1+v_2+v_3\in E_1^x$ we get that 
$v_1+v_2+v_3 \perp v_2$(Condition 3) due to $v_2\in E_2^x$. Similarly
$v_1+v_2 \perp v_3$(Condition 4) and $v_1 \perp v_2$(Condition 2).

It is easy to get that $||P_a(v_1+v_2+v_3)||^2\geq p$(Condition 7) because 
reading $yz$ in the state $q_1$ leads to accepting state.

Let $p_{a_1}$($p_{r_1}$) be 
the accepting(rejecting) probability while reading 
an infinite sequence of letters $y$ in the state $v_1+v_2+v_3$. 
Then $p_{a_1}+p_{r_1}=||v_3||^2$(Condition 5)
due to $v_1+v_2\in E_1^y$ and $v_3\in E_2^y$.

Let $p_{a_2}$($p_{r_2}$) be 
the accepting(rejecting) probability while reading 
an infinite sequence of letters $x$ in the state $v_1+v_2$. 
Then $p_{a_2}+p_{r_2}=||v_2||^2$(Condition 6)
due to $v_1\in E_1^x$ and $v_2\in E_2^x$.

We find an integer $i$ such that after reading $y^i$ 
the norm of $\psi_{\kappa y^i}-(v_1+v_2)$ 
is at most some fixed $\epsilon >0$. 
Now similarly to Theorem \ref{T1}
we can get Condition 8: $||P_a(v_1+v_2)||^2+p_{a_1}\geq p$.



Let $\psi_{\kappa y^i}=\psi_1+\psi_2$, $\psi_1\in E_1^x$, $\psi_2\in E_2^x$.
We find an integer $j$ such that after reading $x^j$ 
the norm of $\psi_{\kappa y^ix^j}-\psi_1$ is at most $\epsilon$.
Since $\psi_1-v_1\perp \psi_2-v_2$ then 
$||\psi_1-v_1||^2+||\psi_2-v_2||^2=||\psi_{\kappa y^i}-(v_1+v_2)||^2<\epsilon^2$.
Therefore, $||\psi_1-v_1||<\epsilon$. 
Then $||\psi_{\kappa y^ix^j}-v_1||\leq ||\psi_{\kappa y^ix^j}-\psi_1||
+||\psi_1-v_1||<2\epsilon$ due to previous inequalities. 
Now similarly to Theorem \ref{T1}
we can get Condition 9: $||P_a(v_1)||^2+p_{a_1}+p_{a_2}\leq 1-p$.\\

We have constructed our second optimization problem. 
We solve the problem by computer. 
Using this solution we can easily construct corresponding 
quantum automaton.
\qed

%

\subsection{$k$ cycles in parallel}

\begin{figure}
\begin{center}
\epsfxsize=2.5in
\hspace{0in}
\epsfbox{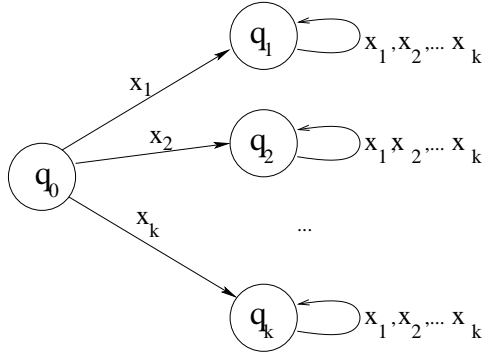}
\caption{\it ``The forbidden construction'' of Theorem \ref{TF5}.}
\label{F5}
\end{center}
\end{figure}

\begin{thm}
\label{TF5}
Let $k\geq 2$.
\begin{enumerate}
\item
Let $L$ be a language. If there are words $x_1, x_2, \ldots, x_k$ 
such that its minimal automaton $M$ contains states 
$q_0, q_1, \ldots, q_k$ satisfying:
\begin{enumerate}
\item
if M starts in the state $q_0$ and reads $x_i$, it passes to $q_i$,
\item
if M starts in the state $q_i(i\geq 1)$ and reads $x_j$, it passes to $q_i$,
\item
for each $i$ the state $q_i$ is not ``all-rejecting'' state,
\end{enumerate}
Then $L$ cannot be recognized by a QFA with probability greater
than $\frac{k}{2k-1}$.
\item
There is a language such that its minimal deterministic automaton 
contains this construction  
and the language can be recognized by a QFA with probability
$\frac{k}{2k-1}$.
\end{enumerate}
\end{thm}

For $k=2$, a related construction was considered in \cite{AKV 01}.
There is a subtle difference between the two constructions (the one considered
here for $k=2$ and the one in \cite{AKV 01}).
The ``non-reversible construction'' in \cite{AKV 01} 
requires the sets of words accepted 
from $q_1$ and $q_2$ to be incomparable.
This extra requirement makes it much harder:
no QFA can recognize a language with the ``non-reversible construction'' of 
\cite{AKV 01} even with the probability $1/2+\epsilon$. 

{\em Proof.}

{\bf Impossibility result.}
This is the only proof in this paper that does not use
a reduction to an optimization problem.
Instead, we use a variant of the classification 
of states (Lemma \ref{AKVLemma}) directly.

We only consider the case when 
the sets of words accepted from 
$q_i$ and $q_j$ are not incomparable.
(The other case follows from the impossibility result in
\cite{AKV 01}.)
                                            
Let $L_i$ be the set of words accepted from $q_i(i\geq 1)$. 
This means that for each $i,j$ we have 
either $L_i\subset L_j$ or $L_j\subset L_i$. 
Without loss of generality we can assume that 
$L_1\subset L_2\subset \ldots \subset L_k$. 
Now we can choose $k$ words $z_1, z_2, \ldots, z_k$ such that 
$z_i\in L_1, L_2, \ldots, L_{k+1-i}$ and $z_i\notin L_{k+2-i}, \ldots, L_k$.
The word $z_1$ exists due to the condition {\it (c)}.

We use a generalization of Lemma \ref{AKVLemma}.

\begin{lem}
\label{Lemma2}
Let $x_1,\ldots, x_k\in \Sigma^{+}$.
There are subspaces $E_1$, $E_2$ such that $E_{non}=E_1\oplus E_2$ and
\begin{enumerate}
\item[(i)]
If $\psi\in E_1$, then 
$V'_{x_1}(\psi)\in E_1,$
$\ldots,$
$V'_{x_k}(\psi)\in E_1$
and
$\|V'_{x_1}(\psi)\|=\|\psi\|,$
$\ldots,$
$\|V'_{x_k}(\psi)\|=\|\psi\|,$
\item[(ii)]
If $\psi\in E_2$, then for any $\epsilon >0$,
there exists a word $t\in (x_1|\ldots|x_k)^*$ such that
$\| V'_t(\psi)\|<\epsilon$.
\end{enumerate}
\end{lem}

The proof is similar to lemma \ref{AKVLemma}.

Let $L$ be a language such that its minimal automaton $M$
contains the ''non reversible construction'' from Theorem \ref{TF5}
and $M_q$ be a QFA. Let $p$ be the accepting probability of $M_q$.
We show that $p\leq \frac{k}{2k-1}$. 

Let $w$ be a word such that after reading it $M$
is in the state $q_0$.
Let $\psi_{w}=\psi^1_{w}+\psi^2_{w}$,
$\psi^1_{w}\in E_1$, $\psi^2_{w}\in E_2$.
We find a word $a_1\in (x_1|\ldots |x_k)^*$ such that after reading $x_1a_1$ 
the norm of 
$\psi^2_{wx_1a_1}=V'_{a_1}(\psi^2_{wx_1})$ is at most 
some fixed $\epsilon>0$.
(Such word exists due to Lemma \ref{Lemma2}.)
We also find words $a_2,\ldots, a_k$ such that
$\|\psi^2_{wx_2a_2}\|\leq \epsilon$, $\ldots$,
$\|\psi^2_{wx_ka_k}\|\leq \epsilon$.

Because of unitarity of $V'_{x_1}$, $\ldots$, $V'_{x_k}$ on $E_1$ 
(part (i) of Lemma \ref{Lemma2}),
there exist integers $i_1\ldots i_k$ such that
$\|\psi^1_{w(x_1a_1)^{i_1}}-\psi^1_{w}\|\leq\epsilon$,
$\ldots,$
$\|\psi^1_{w(x_ka_k)^{i_k}}-\psi^1_{w}\|\leq\epsilon$.

Let $p_w$ be the probability of $M_q$ accepting while reading $\kappa w$.
Let $p_1,\ldots, p_k$ be the probabilities of accepting while reading 
$(x_1a_1)^{i_1},\ldots, (x_ka_k)^{i_k}$ with a starting state $\psi_w$ and 
and $p^{'}_1,\ldots, p^{'}_k$ be
the probabilities of accepting while reading $z_1\$,\ldots, z_k\$$
with a starting state $\psi^1_w$.

Let us consider $2k-1$ words:\\
$\kappa w (x_1a_1)^{i_1}z_k\$,$\\
$\kappa w (x_2a_2)^{i_2}z_k\$,$\\
$\kappa w (x_2a_2)^{i_2}z_{k-1}\$,$\\
$\kappa w (x_3a_3)^{i_3}z_{k-1}\$,$\\
$\ldots,$\\
$\kappa w (x_{k-1}a_{k-1})^{i_{k-1}}z_2\$,$\\
$\kappa w (x_ka_k)^{i_k}z_2\$,$\\
$\kappa w (x_ka_k)^{i_k}z_1\$.$

\begin{lem}
$M_q$ accepts $\kappa w (x_1a_1)^{i_1}z_k\$$
with probability at least $p_w+p_1+p^{'}_k-4\epsilon$
and at most $p_w+p_1+p^{'}_k+4\epsilon$.
\end{lem}

{\bf Proof.}
The probability of accepting while reading $\kappa w$ is $p_w$.
After that, $M_q$ is in the state $\psi_w$ and reading $(x_1a_1)^{i_1}$
in this state causes it to accept with probability $p_1$.

The remaining state is $\psi_{w(x_1a_1)^{i_1}}=
\psi^1_{w(x_1a_1)^{i_1}}+\psi^2_{w(x_1a_1)^{i_1}}$.
If it was $\psi^1_w$, the probability of accepting while
reading the rest of the word ($z_k\$$) would be exactly $p^{'}_k$.
It is not quite $\psi^1_w$ but it is close to $\psi^1_w$.
Namely, we have 
\[ \|\psi_{w(x_1a_1)^{i_1}}-\psi^1_w\| \leq \|\psi^2_{w(x_1a_1)^{i_1}}\|
+\|\psi^1_{w(x_1a_1)^{i_1}}-\psi^1_w\| \leq \epsilon+\epsilon =2\epsilon.\]
By Lemma \ref{BVLemma}, this means that the probability of accepting
during $z_k\$$ is between $p^{'}_k-4\epsilon$ and $p^{'}_k+4\epsilon$.
\qed

This Lemma implies that $p_w+p_1+p^{'}_k+4\epsilon\geq p$ because 
of $x_1z_k\in L$.
Similarly, $1-p_w-p_2-p^{'}_k+4\epsilon\geq p$ because of $x_2z_k\notin L$.
Finally, we have $2k-1$ inequalities:\\
$p_w+p_1+p^{'}_k+4\epsilon\geq p,$\\
$1-p_w-p_2-p^{'}_k+4\epsilon\geq p,$\\
$p_w+p_2+p^{'}_{k-1}+4\epsilon\geq p,$\\
$1-p_w-p_3-p^{'}_{k-1}+4\epsilon\geq p,$\\
$\ldots,$\\
$p_w+p_{k-1}+p^{'}_2+4\epsilon\geq p,$\\
$1-p_w-p_k-p^{'}_2+4\epsilon\geq p,$\\
$p_w+p_k+p^{'}_1+4\epsilon\geq p.$\\

By adding up these inequalities we get 
$k-1+p_w+p_1+p^{'}_1+4(2k-1)\epsilon\geq (2k-1)p$.
We can notice that $p_w+p_1+p^{'}_1\leq 1$.
(This is due to the facts that 
$p_1\leq ||\psi^2_{w}||^2$, $p^{'}_1\leq ||\psi^1_{w}||^2$
and $1-p_w\leq||\psi_{w}||^2=||\psi^2_{w}||^2+||\psi^1_{w}||^2$.)
Hence, $p\leq \frac{k}{2k-1}+4\epsilon$. Since such $2k-1$
words can be constructed for arbitrarily small $\epsilon$, this means 
that $M_q$ does not recognize $L$ with probability greater than 
$\frac{k}{2k-1}$. \qed

{\bf Constructing a quantum automaton.}

We consider a language $L_1$ in the alphabet $b_1,b_2, \ldots, b_k,
z_1, z_2, \ldots, z_k$ such that its minimal automaton
has accepting states $q_0, q_1, \ldots, q_k$ and rejecting state
$q_{rej}$ and the 
transition function $V_1$ is defined as follows:

$V_1(q_0,b_i)=q_i$, $V_1(q_0,z_i)=q_1$,
$V_1(q_i,b_j)=q_i(i>1)$, 
$V_1(q_i,z_j)=q_1(i+j\leq k+1)$,
$V_1(q_i,z_j)=q_{rej}(i+j> k+1)$,
$V_1(q_{rej},b_i)=q_{rej}$, $V_1(q_{rej},z_i)=q_{rej})$.

It can be checked that this automaton contains the 
''non reversible construction'' from Theorem 4. 
Hence, this language cannot be recognized by a QFA with probability greater
than $\frac{k}{2k-1}$.

Next, we construct a QFA $M_q$ that accepts this language with 
such probability.

The automaton has $3(k+1)$ states: 
$q^{'}_0, q^{'}_2,\ldots, q^{'}_k$,
$q_{a_0}, q_{a_2},\ldots, q_{a_k}$,
$q_{r_0}, q_{r_2},\ldots,$ $q_{r_k}$.
$Q_{acc}=\{ q_{a_0}, q_{a_2},\ldots, q_{a_k}\}$, 
$Q_{rej}=\{ q_{r_0}, q_{r_2},\ldots, q_{r_k}\}$.
The initial state is 
$$\sqrt{\frac{k}{2k-1}}|q^{'}_0\rangle+
  \sqrt{\frac{1}{2k-1}}|q^{'}_2\rangle+\ldots
  \sqrt{\frac{1}{2k-1}}|q^{'}_k\rangle.$$ 
The transition function is 
$$V_{b_i}(|q^{'}_0\rangle)=\sqrt{\frac{k+1-i}{k}}|q_{a_0}\rangle+\sqrt{\frac{i-1}{k}}
|q_{r_0}\rangle,
V_{b_i}(|q^{'}_j\rangle)=|q^{'}_j\rangle(j\geq 2),$$
$$V_{z_i}(|q^{'}_0\rangle)=|q_{a_0}\rangle,
V_{z_i}(|q^{'}_j\rangle)=|q_{a_j}\rangle(i+j\leq k+1),
V_{z_i}(|q^{'}_j\rangle)=|q_{r_j}\rangle(i+j> k+1),$$
$$V_\$(|q^{'}_j\rangle)=|q_{a_j}\rangle.$$

\begin{enumerate}
\item
The empty word.

The only tranformation applied to the starting state is $V_{\$}$.
Therefore, the final superposition is
$$\sqrt{\frac{k}{2k-1}}|q_{a_0}\rangle+
  \sqrt{\frac{1}{2k-1}}|q_{a_2}\rangle+\ldots
  \sqrt{\frac{1}{2k-1}}|q_{a_k}\rangle$$ 
and the word is accepted with probability 1. 

\item
The word starts with $z_i$.

Reading $z_i$ maps $|q^{'}_0\rangle$ to $|q_{a_0}\rangle$.
Therefore, 
this word is accepted with probability at least
$(\sqrt{\frac{k}{2k-1}})^2=\frac{k}{2k-1}$.

\item
Word is in form $b_i(b_1\vee \ldots \vee b_k)^*$.
The superposition after reading $b_i$ is 
$$
\sqrt{\frac{k+1-i}{2k-1}}|q_{a_0}\rangle+
\sqrt{\frac{i-1}{2k-1}}|q_{r_0}\rangle+
\sqrt{\frac{1}{2k-1}}|q^{'}_2\rangle+\ldots
\sqrt{\frac{1}{2k-1}}|q^{'}_k\rangle.
$$
At this moment $M_q$ accepts with probability $\frac{k+1-i}{2k-1}$  
and rejects with probability $\frac{i-1}{2k-1}$. The computation
continues in the superposition 
$$\sqrt{\frac{1}{2k-1}}|q^{'}_2\rangle+\ldots
\sqrt{\frac{1}{2k-1}}|q^{'}_k\rangle.$$
Clearly, that reading of all remaining letters does not change this 
superposition. Since $V_\$$ maps each $|q^{'}_j\rangle$ to an accepting state
then $M_q$ rejects this word with probability at most 
$\frac{i-1}{2k-1}\leq \frac{k-1}{2k-1}.$

\item
Word $x$ starts with $b_i(b_1\vee \ldots \vee b_k)^*z_j$.
Before reading $z_j$ the superposition is
$$\sqrt{\frac{1}{2k-1}}|q^{'}_2\rangle+\ldots
\sqrt{\frac{1}{2k-1}}|q^{'}_k\rangle.$$

{\it Case 1.} $i+j>k+1$. $x\notin L_1.$\\
Since $i+j>k+1$ then reading $z_j$ maps at least $k-i+1$ states of 
$q^{'}_2,\ldots ,q^{'}_k$ to rejecting states. 
This means that $M_q$ rejects with probability at least 
$$\frac{i-1}{2k-1}+\frac{k-i+1}{2k-1}=\frac{k}{2k-1}.$$

{\it Case 2.} $i+j\leq k+1$. $x\in L_1.$
Since $i+j\leq k+1$ then reading $z_j$ maps at least $i-1$ states of 
$q^{'}_2,\ldots ,q^{'}_k$ to accepting states. 
This means that $M_q$ accepts with probability at least 
$$\frac{k+1-i}{2k-1}+\frac{i-1}{2k-1}=\frac{k}{2k-1}.$$
\end{enumerate}
\qed


\subsection{$0.7324...$ construction}

\begin{thm}
\label{TF2}
Let $L$ be a language. 
\begin{enumerate}
\item
If there are words $x$, $z_1$, $z_2$ such that its minimal automaton $M$ 
contains states $q_1$ and $q_2$ satisfying:
\begin{enumerate}
\item
if M starts in the state $q_1$ and reads $x$, it passes to $q_2$,
\item
if M starts in the state $q_2$ and reads $x$, it passes to $q_2$,
\item
if M starts in the state $q_1$ and reads $z_1$, it passes to
an accepting state,
\item
if M starts in the state $q_1$ and reads $z_2$, it passes to
a rejecting state,
\item
if M starts in the state $q_2$ and reads $z_1$, it passes to
a rejecting state,
\item
if M starts in the state $q_2$ and reads $z_2$, it passes to
an accepting state.
\end{enumerate}
Then $L$ cannot be recognized by a QFA with probability greater
than $\frac{1}{2}+\frac{3\sqrt{15}}{50}=0.7324...$.
\item
There is a language $L$ with the minimum automaton containing
this construction that can be recognized with probability
$\frac{1}{2}+\frac{3\sqrt{15}}{50}=0.7324...$.
\end{enumerate}
\end{thm}

\begin{figure}
\begin{center}
\epsfxsize=3in
\hspace{0in}
\epsfbox{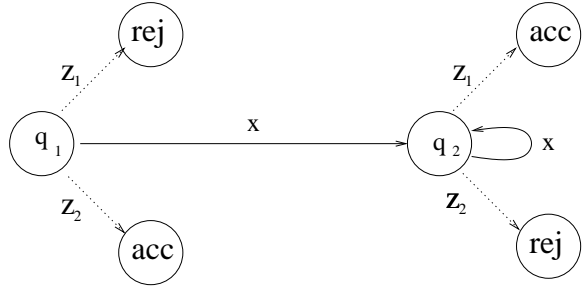}
\caption{\it ``The forbidden construction'' of Theorem \ref{TF2}.}
\label{F2}
\end{center}
\end{figure}

\noindent
{\em Proof.}

{\bf Impossibility result.}

The construction of optimization problem is similar to 
the construction of Optimization problem 1.
For this reason, we omit it and just give the optimization
problem and show how to solve it.


{\bf Optimization problem 3.}
Find the maximum $p$ such that there is a finite dimensional
vector space $E_{opt}$, subspaces $E_a$, $E_r$ 
(unlike in previous optimization problems, $E_a$ and $E_r$ do not
have to be orthogonal)
and vectors $v_1$, $v_2$ such that $v_1\perp v_2$ and $\|v_1+v_2\|=1$
and probabilities $p_1$, $p_2$ such that $p_1+p_2=\|v_2\|^2$ and
\begin{enumerate}
\item
$\|P_a(v_1+v_2)\|^2\geq p$,
\item
$\|P_r(v_1+v_2)\|^2\geq p$,
\item
$1-\|P_a(v_1)\|^2-p_1\geq p$,
\item
$1-\|P_r(v_1)\|^2-p_2\geq p$.
\end{enumerate}

{\bf Solving optimization problem 3.}

Without loss of generality we can assume that
$\|P_a(v_1)\|\leq\|P_r(v_1)\|$. Then these four 
inequalities can be replaced
with only three inequalities
\begin{enumerate}
\item
$\|P_a(v_1+v_2)\|^2\geq p$,
\item
$1-\|P_a(v_1)\|^2-p_1\geq p$,
\item
$1-\|P_a(v_1)\|^2-p_2\geq p$.
\end{enumerate}
Clearly that $p$ is maximized by $p_1=p_2=\frac{\|v_2\|^2}{2}$. Therefore, 
we have 
\begin{enumerate}
\item
$\|P_a(v_1+v_2)\|^2\geq p$,
\item
$1-\|P_a(v_1)\|^2-\frac{\|v_2\|^2}{2}\geq p$.
\end{enumerate}

Next we show that it is enough to consider only instances
of small dimension.
We denote $E_{opt}-E_a$ as $E_b$.
First, we restrict $E_a$ to the subspace $E'_a$ generated by
projections of $v_1$ and $v_2$ to $E_a$. This subspace is
at most 2-dimensional.
Similarly, we restrict $E_b$ to the subspace $E'_b$ generated by 
projections of $v_1$ and $v_2$ to $E_b$.
The lengths of all projections are still the same.
We fix an orthonormal basis for $E_{opt}$ so that $P_a(v_1)$ and
$P_b(v_1)$
are both parallel to some basis vectors.
Then, $v_1=(x_1, 0, x_3, 0)$ and $v_2=(y_1, y_2, y_3, y_4)$ 
where the first two coordinates correspond to basis
vectors of $E'_a$ and the last two coordinates correspond
to basis vectors of $E'_b$.
We can assume that $x_1$ and $x_3$ are both non-negative.
(Otherwise, just invert the direction of one of basis vectors.)

Let $\Delta=\|v_1\|=\sqrt{x_1^2+x_3^2}$.
Then, there is $\alpha\in[0,\pi/2]$ such that $x_1=\Delta\cos\alpha$,
$x_3=\Delta\sin\alpha$. 
Let $\delta=\sqrt{y_1^2+y_3^2}$.
Then, $y_1=\delta\sin\alpha$, $y_3=-\delta\cos\alpha$
because $x_1 y_1+x_3 y_3=0$ due to $v_1\perp v_2$.
If $y_4\neq 0$, we can change $y_1$ and $y_3$ to
$\delta'\sin\alpha$ and $-\delta'\cos\alpha$
where $\delta'=\sqrt{y_1^2+y_3^2+y_4^2}$ and
this only increases $\|P_a(v_1+v_2)\|$.
Hence, we can assume that $y_4=0$.
We denote $\epsilon=y_2$. Then, 
$v_1=(\Delta\cos\alpha, 0, \Delta\sin\alpha, 0)$,
$v_2=(\delta\sin\alpha, \epsilon, -\delta\cos\alpha, 0)$.

Let ${\rm E}=\sqrt{\Delta^2+\delta^2}$. Then,
$\Delta={\rm E}\sin\beta$ and $\delta={\rm E}\cos\beta$
for some $\beta\in[0, \pi/2]$ and $E^2+\epsilon^2=1$. This gives
\begin{enumerate}
\item
$\|P_a(v_1+v_2)\|^2=E^2(\sin\beta\cos\alpha+\cos\beta\sin\alpha)^2+
\epsilon^2=
E^2\sin^2(\alpha+\beta)+\epsilon^2\geq p$,
\item
$1-\|P_a(v_1)\|^2-\frac{\|v_2\|^2}{2}=
1-E^2\sin^2\beta\cos^2\alpha-\frac{E^2\cos^2\beta+\epsilon^2}{2}\geq p$.
\end{enumerate}
Then after some calculations we get
\begin{enumerate}
\item
$1-E^2\cos^2(\alpha+\beta)\geq p$,
\item
$\frac{1-E^2\sin^2\beta\cos2\alpha}{2}\geq p$.
\end{enumerate}
If we fix $\alpha+\beta$ and vary $\beta$,
then $-\sin^2\beta \cos2\alpha$ (and, 
hence, 
$\frac{1-E^2\sin^2\beta\cos2\alpha}{2}$)
is maximized by $\beta=2\alpha-\pi/2$. This means that
we can assume $\beta=2\alpha-\pi/2$ and we have
\begin{enumerate}
\item
$1-E^2\sin^2(3\alpha)\geq p$,
\item
$\frac{1-E^2\cos^3(2\alpha)}{2}\geq p$.
\end{enumerate}
If we consider $\cos^2\alpha\geq 1/2$ then
$p\leq\frac{1-E^2\cos^3(2\alpha)}{2}=
\frac{1-E^2(2\cos^2\alpha-1)^3}{2}\leq 1/2$. This means that we are only 
interested in $\cos^2\alpha< 1/2$.

Let $f(E^2,\alpha)=1-E^2\sin^2(3\alpha)$ and
$g(E^2,\alpha)=\frac{1-E^2\cos^3(2\alpha)}{2}$.
If we fix $\alpha$ and vary $E^2$, then $f$ and $g$ are 
linear functions in $E^2$ and $f(0,\alpha)>g(0,\alpha)$.
We consider two cases. \\
\\
{\it Case 1.} $f(1,\alpha)\geq g(1,\alpha)$. (This gives 
$f(E^2,\alpha)\geq g(E^2,\alpha)$ for each $E^2$. Therefore, in this case 
we only need to maximize the function $g$.)\\
This means that 
$$1-\sin^2(3\alpha)\geq\frac{1-\cos^3(2\alpha)}{2},$$
$$1-2\sin^2(3\alpha)+\cos^3(2\alpha)\geq 0,$$
$$1-2(1-\cos^2(3\alpha))+\cos^3(2\alpha)\geq 0,$$
$$1-2(1-(4\cos^3\alpha-3\cos\alpha)^2)+\cos^3(2\alpha)\geq 0,$$
$$1-2(1-16\cos^6\alpha+24\cos^4\alpha-9\cos^2\alpha)+(2\cos^2\alpha-1)^3\geq
0,$$
$$20\cos^6\alpha-30\cos^4\alpha+12\cos^2\alpha-1\geq 0,$$
$$(1-2\cos^2\alpha)(-10\cos^4\alpha+10\cos^2\alpha-1)\geq 0.$$
So that $\cos^2\alpha< 1/2$, we have
$$-10\cos^4\alpha+10\cos^2\alpha-1\geq 0.$$
This means that
$\cos^2\alpha\in[\frac{1}{2}-\frac{\sqrt{15}}{10},\frac{1}{2}]$.

Since 
$g(E^2,\alpha)=\frac{1-E^2(2\cos^2\alpha-1)^3}{2}$, $g$ is
maximized by $E^2=1$ and $\cos^2\alpha=\frac{1}{2}-\frac{\sqrt{15}}{10}$.
This gives $p$ equal to $\frac{1}{2}+\frac{3\sqrt{15}}{50}$.\\

{\it Case 2.} $f(1,\alpha)\leq g(1,\alpha)$.
(This is equivalent to 
$\cos^2\alpha\in[0,\frac{1}{2}-\frac{\sqrt{15}}{10}]$.)\\
This means that $p$ is maximized by $f(E^2,\alpha)=g(E^2,\alpha)$.
Therefore, 
\begin{enumerate}
\item
$1-E^2\sin^2(3\alpha)=p$,
\item
$\frac{1-E^2\cos^3(2\alpha)}{2}=p$.
\end{enumerate}
Let $y$ be $-\cos 2\alpha=1-2\cos^2\alpha$.
Then $y\in[\sqrt{\frac{3}{5}},1]$ and
$\sin^2(3\alpha)=1-\cos^2(3\alpha)=1-(4\cos^3\alpha-3\cos\alpha)^2=
1-\cos^2\alpha(4\cos^2\alpha-3)^2=1-\frac{1-y}{2}(1+2y)^2=\frac{1-3y+4y^3}{2}$.
Therefore,
\begin{enumerate}
\item
$2-E^2(4y^3-3y+1)=2p$,
\item
$1+E^2y^3=2p$.
\end{enumerate}
Now we express $p$ using only $y$. 
We get $p=\frac{1}{2}+\frac{y^3}{2(5y^3-3y+1)}$. Finally, if we vary
$y$ through the interval $[\sqrt{\frac{3}{5}},1]$, then 
$p$ is maximized by $y=\sqrt{\frac{3}{5}}$. This gives $p$
equal to $\frac{1}{2}+\frac{3\sqrt{15}}{50}$.  \qed

{\bf Construction of a QFA.}

We consider the two letter alphabet $\{a,b\}$. The language $L$ is the 
union of the empty word and $a^+b(a\vee b)^*$. 
Clearly that the minimal deterministic automaton of $L$ contains
the ''non reversible construction'' from Theorem 5
(just take $a$ as $x$, the empty word 
as $z_1$ and $b$ as $z_2$).

Next, we describe a QFA $M$ accepting this language. 
Let $\alpha$ be the solution of
$1-2\cos^2\alpha=\sqrt{\frac{3}{5}}$ in the interval $[0,\pi/2]$.
It can be checked that $\cos^2(3\alpha)=\frac{1}{2}+\frac{3\sqrt{15}}{50}$,
$\sin^22\alpha=\frac{2}{5}$, 
$\cos^22\alpha=\frac{3}{5}$, 
$\sin^2\alpha=\frac{1}{2}+\frac{\sqrt{3}}{2\sqrt{5}}$.

The automaton has 4 states: $ q_0,  q_1,  q_{acc}$ and $ q_{rej}$.
$Q_{acc}=\{ q_{acc}\}$, $Q_{rej}=\{ q_{rej}\}$.
The initial state is $\cos(3\alpha) |q_0\rangle+\sin(3\alpha) |q_1\rangle$. 
The transition function is 

\[ V_a(|q_0\rangle)=\cos^2\alpha|q_0\rangle+
\cos\alpha\sin\alpha|q_{1}\rangle+
\frac{\sin\alpha}{\sqrt{2}}|q_{acc}\rangle+
\frac{\sin\alpha}{\sqrt{2}}|q_{rej}\rangle,\]
\[ V_a(|q_1\rangle)=\cos\alpha\sin\alpha|q_0\rangle+
\sin^2\alpha|q_{1}\rangle-
\frac{\cos\alpha}{\sqrt{2}}|q_{acc}\rangle-
\frac{\cos\alpha}{\sqrt{2}}|q_{rej}\rangle,\]
\[ V_b(|q_0\rangle)=|q_{rej}\rangle, 
V_b(|q_1\rangle)=|q_{acc}\rangle,\]
\[ V_{\$}(|q_0\rangle)=|q_{acc}\rangle,
 V_{\$}(|q_1\rangle)=|q_{rej}\rangle,\]

\begin{enumerate}
\item
The empty word.

The only tranformation applied to the starting state is $V_{\$}$.
Therefore, the final superposition is
$\cos(3\alpha) |q_{acc}\rangle+\sin(3\alpha) |q_{rej}\rangle$ and 
the word is accepted with probability
$\cos^2(3\alpha)=\frac{1}{2}+\frac{3\sqrt{15}}{50}$. 

\item
$b(a\vee b)^*$.

After reading $b$ the superposition is 
$\sin(3\alpha) |q_{acc}\rangle+\cos(3\alpha) |q_{rej}\rangle$ and 
word is rejected with probability
$\cos^2(3\alpha)=\frac{1}{2}+\frac{3\sqrt{15}}{50}$. 

\item
$a^+$.

After reading the first $a$ the superposition becomes
$$\cos\alpha\cos2\alpha |q_0\rangle+
\sin\alpha\cos2\alpha |q_1\rangle-
\frac{\sin2\alpha}{\sqrt{2}} |q_{acc}\rangle-
\frac{\sin2\alpha}{\sqrt{2}} |q_{rej}\rangle.$$
At this moment $M$ accepts with probability
$\frac{\sin^22\alpha}{2}=\frac{1}{5}$ and rejects with
probability $\frac{1}{5}$.
The computation continues in the superposition
$$\cos\alpha\cos2\alpha |q_0\rangle
+\sin\alpha\cos2\alpha |q_1\rangle.$$
It is easy to see that reading all of remaining letters
does not change this superposition. 

Therefore, the final superposition (after reading $\$$) is
$$\cos\alpha\cos2\alpha |q_{acc}\rangle+
\sin\alpha\cos2\alpha |q_{rej}\rangle.$$
This means that $M$ rejects with probability
$$\sin^2\alpha\cos^22\alpha+\frac{1}{5}=
\frac{3}{5}(\frac{1}{2}+\frac{\sqrt{3}}{2\sqrt{5}})+\frac{1}{5}=
\frac{1}{2}+\frac{3\sqrt{15}}{50}$$.

\item
$a^+b(a\vee b)^*$.

Before reading the first $b$ the superposition is
$$\cos\alpha\cos2\alpha |q_0\rangle
+\sin\alpha\cos2\alpha |q_1\rangle$$ and
reading this $b$ changes this superposition to
$$\sin\alpha\cos2\alpha |q_{acc}\rangle
+\cos\alpha\cos2\alpha |q_{rej}\rangle.$$ 
This means that $M$ accepts with probability
$$\sin^2\alpha\cos^22\alpha
+\frac{1}{5}=
\frac{1}{2}+\frac{3\sqrt{15}}{50}.$$
\end{enumerate}
\qed

\section{Conclusion}

Quantum finite automata (QFA) can recognize all regular languages if 
arbitrary intermediate measurements are allowed.
If they are restricted to be unitary, the computational power drops
dramatically, to languages recognizable by permutation automata 
\cite{CM 97,BP 99}.
In this paper, we studied an intermediate case in which measurements
are allowed but restricted to "accept-reject-continue" form 
(as in \cite{KW 97,AF 98,BP 99}).

Quantum automata of this type can recognize several languages
not recognizable by the corresponding classical model (reversible finite automata).
In all of those cases, those languages cannot be recognized with
probability 1 or $1-\epsilon$, but can be recognized with some fixed probability $p>1/2$.
This is an unusual feature of this model because, in most other computational 
models a probability of correct answer $p>1/2$ can be easily amplified
to $1-\epsilon$ for arbitrary $\epsilon>0$.

In this paper, we study maximal probabilities of correct answer achievable
for several languages. Those probabilities are related to ``forbidden constructions"
in the minimal automaton. A ``forbidden construction" being present in the 
minimal automaton implies that the language cannot be recognized with 
a probability higher than a certain $p>1/2$. 

The basic construction is ``one cycle" in figure \ref{F1}.
Composing it with itself sequentially (figure \ref{F4})
or in parallel (figure \ref{F5}) gives ``forbidden constructions"
with a smaller probability $p$.
The achievable probability also depends on whether 
the sets of words accepted from the different states 
of the construction are subsets of one another (as in figure \ref{F1})
or incomparable (as in figure \ref{F2}). 
The constructions with incomparable sets usually imply
smaller probabilities $p$.

The accepting probabilities $p$ quantify the degree of non-reversibility
present in the ``forbidden construction". 
Lower probability $p$ means that the language is more difficult for
QFA and thus, the ``construction" has higher degree of non-reversibility.
In our paper, we gave a method for calculating this probability
and used it to calculate the probabilities $p$ for several ``constructions".
The method should apply to a wide class of constructions but solving
the optimization problems can become difficult if the 
construction contains more states (as for language $a_1^* a_2^* \ldots a^*_k$
studied in \cite{ABFK 99}).
In this case, it would be good to have methods
for calculating the accepting probabilities approximately.

A more general problem suggested by this work is: how do we 
quantify non-reversibility?
Accepting probabilities of QFAs provide one way of 
comparing the degree of non-reversibility 
in different ``constructions''. 
What are the other ways of quantifying it?
And what are the other settings in which 
similar questions can be studied?

\end{document}